# Low Resistivity and High Breakdown Current Density of 10-nm Diameter van der Waals TaSe$_3$ Nanowires by Chemical Vapor Deposition

Thomas A. Empante[1], Aimee Martinez[1], Michelle Wurch[1], Yanbing Zhu[2], Adane K. Geremew[3], Koichi Yamaguchi[1], Miguel Isarraraz[1], Sergey Rumyantsev[3,4], Evan J. Reed[2], Alexander A. Balandin[3], Ludwig Bartels[1]

[1]Department of Chemistry and Material Science & Engineering Program, University of California – Riverside, Riverside, CA 92521, United States

[2]Department of Materials Science and Engineering, Stanford University, Stanford, California 94304, United States

[3]Nano-Device Laboratory, Department of Electrical and Computer Engineering, University of California, Riverside, California 92521, United States

[4]Center for Terahertz Research and Applications, Institute of High Pressure Physics, Polish Academy of Sciences, Warsaw 01-142, Poland

Abstract:

**Micron-scale single-crystal nanowires of metallic TaSe$_3$, a material that forms -Ta-Se$_3$-Ta-Se$_3$- stacks separated from one another by a tubular van der Waals (vdW) gap, have been synthesized using chemical vapor deposition (CVD) on a SiO$_2$/Si substrate, in a process compatible with semiconductor industry requirements. Their electrical resistivity was found unaffected by downscaling from the bulk to as little as 7 nm in width and height, in striking contrast to the resistivity of copper for the same dimensions. While the bulk resistivity of TaSe$_3$ is substantially higher than that of bulk copper, at the nanometer scale the TaSe$_3$ wires become competitive to similar-sized copper ones. Moreover, we find that the vdW TaSe$_3$ nanowires sustain current densities in excess of $10^8$ A/cm$^2$ and feature an electromigration energy barrier twice that of copper. The results highlight the promise of quasi-one-dimensional transition metal trichalcogenides for electronic interconnect applications and the potential of van der Waals materials for downscaled electronics.**

The preparation, characterization, and application of two-dimensional (2D) materials such as graphene[1-4], transition metal dichalcogenides (TMDs)[5-10], and MXenes[11-14] have attracted broad attention over the past years. This is partly due to the unique properties such materials exhibit when thinned to a single layer, yet it is also due to the astounding chemical stability and well-defined optical and electronic properties that these materials retain, even when only a single layer thin. Arguably, the latter is their most outstanding feature; it is caused by the presence of a planar van der Waals (vdW) gap in the crystallographic structure that affords some independence between the layers even in the bulk. When thinned to a single layer, the vdW gap prevents dangling bonds on the basal plane reducing scattering in transport and chemical reactivity to the environment. Recently, some of us have discovered almost 500 known compounds that resemble these 2D materials in as much as that their bulk contains vdW gaps, but these gaps are tubular in nature creating one-dimensional (1D) stacks of bound atoms separated from the neighboring stack similar to how a graphene or $MoS_2$ sheet is separated from the one above and below. Initial transport measurements on metallic 1D vdW tantalum triselenide ($TaSe_3$)[15-17] and zirconium tritelluride ($ZrTe_3$)[18] by some of us revealed good conductivity and exceptionally high breakdown current on mesoscopic exfoliated wire bundles. Here we show that chemical vapor deposition (CVD) allows the fabrication of wire bundles as small as a few nanometers across (i.e. consisting of a hundred atom stacks or less in parallel), and that such nanoscale bundles retain the bulk conductivity, much different to conventional metals, which at the 10 nm cross section scale are strongly affected by surface and grain-boundary scattering.[19-21] Additionally, we find that $TaSe_3$ exhibits a barrier to electromigration more than twice that of copper and can sustain current densities in excess of $10^8$ A/cm$^2$. The favorable scaling of the conductivity with wire cross section renders these material of great interest as next generation interconnect material as copper reaches its scaling limits in 2023 according to the ITRS roadmap.[22-25]

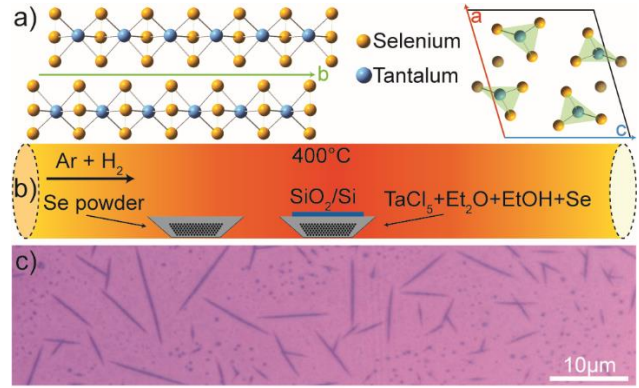

*Fig. 1* a) Crystallographic structure of TaSe$_3$ consisting of four nanowire Ta-Se$_3$ stacks per unit cell along the b axis. The greyed atoms inside the unit cell correspond to those shown outside of it in order to highlight the selenium triangles between each tantalum plane; b) schematic representation of the chemical vapor deposition process inside a tube furnace; c) optical image of a population of TaSe$_3$ nanowires.

This manuscript focuses on the preparation and characterization of a specific transition metal trichalcogenide (TMT), TaSe$_3$, on a commercial 300 nm SiO$_2$/Si substrate using process parameters (ambient pressure, ≤ 400°C process temperature, ≤ 5 min process duration) that are amenable to conventional back-end-of-the-line (BEOL) process limits. Despite the large interest in TMDs, TMTs have been largely left unstudied at the nanoscale; they have varying structures from 2D thin films to 1D wires and properties ranging from insulating to metallic.[17, 26-28] TaSe$_3$ has a monoclinic unit cell and consists of stacks of Ta atoms, each of which are bonded to three selenium atoms above and below along the b axis (Fig. 1a). Neighboring -Ta-Se$_3$-Ta-Se$_3$- stacks are separated from one another by a tubular vdW gap exposing chalcogen atoms only, similar to the gap between the Se-Mo-Se layers in 2D MoSe$_2$. TaSe$_3$ has been known for a long time and bulk samples have been prepared using chemical vapor transport (CVT), a process that requires long process times and is not amenable to current semiconductor processing paradigms; exfoliation of such samples yielded the results reported by some of us earlier.[15-17]

Interconnect performance is crucial for low-power high-clock-frequency computing: the transition from aluminum to copper interconnects starting some 20 years ago was driven by both the better conductivity of copper and the better manageability of electromigration in copper, the key failure mechanism for interconnects.[29, 30] However, as the cross section of an interconnect becomes shorter than the electron mean free path (~40 nm in copper),[19-21] its resistivity increases dramatically due to scattering at the material surface and at internal grain boundaries. A 1D material without surface dangling bonds or internal grain boundaries would, in theory, lack these drawbacks and be a prime candidate.

The best aspect ratio of interconnect cross sections has been studied intensely optimizing interconnect topology while reducing capacitive cross coupling. Modern processors use aspect ratios between 1.2 and 1.5 on the first four (0-3) metal layers.[31] We show a CVD method that natively generates nanoscale wires with a width to height aspect ratio of ~1, close to the optimal one.

The 1D vdW TaSe$_3$ nanowire bundles are synthesized in a chemical vapor deposition (CVD) process using TaCl$_5$ and Se powder as reactants (both 99.99%, Sigma Aldrich). Concurrent volatilization of both reactants under growth conditions is tantamount to sufficient chemical potential of each reactant on the surface during nanowire formation and elongation to the desired length. Thus, in order to match the vapor pressure of the tantalum precursor to that of selenium and the optimal growth temperature of TaSe$_3$, we first add diethyl ether to TaCl$_5$ in an inert nitrogen atmosphere to form the well-known metal chloride adduct TaCl$_5$[OEt$_2$].[32] The adduct is dissolved in ethanol, leading possibly to (partial) ligand exchange; a small amount of gas evolution is observed when adding the ethanol. The supporting materials section provide more details on these adducts. Elemental selenium powder suspended in selenium-saturated ethanol is added, and the reactants are well mixed and dried in an alumina crucible. The supporting material section show micrographs that confirm the absence of nanowire formation at this stage. A 300 nm SiO$_2$/Si wafer substrate is placed across the alumina crucible, which is centered in a quartz process tube. A second crucible with elemental selenium is placed upstream so as to maintain selenium pressure during the entire growth process (Fig. 1b). The process tube is placed in a tube furnace; a mixture of argon and hydrogen is used as ambient-pressure process gases. The furnace is heated to 400°C as fast as possible (~ 15 minutes), held for a few minutes (see below), and then rapidly cooled to ambient temperatures resulting in populations of TaSe$_3$ nanowires as shown in Fig. 1 c).

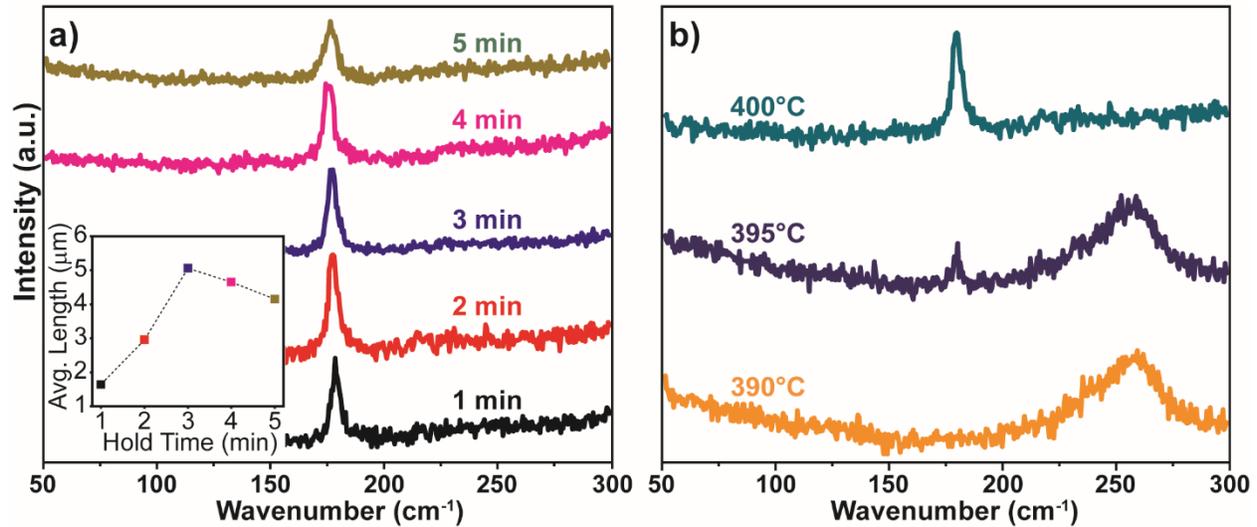

*Fig. 2* : *a) Raman spectra of TaSe$_3$ nanowire bundles held for different duration at the growth temperature of 400°C. The optimal shape of the Raman peak is attained at 2-3 minutes. The inset shows the resulting length distribution peaking at 3 min hold time; b) Nanowire Raman spectra as a function of growth temperature. Starting at 400 °C the desired peak at ~180 cm$^{-1}$ is dominant.*

Raman spectroscopy (Horiba Labram HR800, 532 nm wavelength laser, 0.8 mW of laser power on sample) was used to characterize the nanowire bundles grown at hold times varying from 1 to 5 minutes (Fig. 2a) at 400°C; while nanorods were formed in each instance, 3 minutes hold produced the nanowires with the sharpest Raman signature at ~180 cm$^{-1}$. Determining the average length of a large set of nanowires generated at each hold time, we find a maximum wire length at 3 min hold time (inset in Fig. 2a). This finding suggests that during the hold time the nanowires do not only form and elongate, but also can decomposing, presumably from selenium loss. We studied the Raman spectra of the nanowires as a function of the peak process temperature (Fig. 2b) and find that 400°C is the minimum temperature at which nanowires with the desired Raman signature form. The nanowires with the broad spectral feature at ~260 cm$^{-1}$ formed at process temperatures below 400°C exhibit transport properties far inferior to those nanowires described in the remainder of this manuscript; their precise composition is unknown to us.

In order to be of technological relevance, the TaSe$_3$ 1D vdW nanowires need to be prepared on the scale of a few nanometers in cross section yet significant in length. To this end we evaluated the length to width to height ratios in a population of nanowires. Fig. 3a and b show atomic force and scanning electron microscopy (AFM and SEM, respectively) images of the same population of growth seeds and short wires. We obtain the nanowire height from atomic force microscopy yet we use SEM to establish the nanowire width because of the finite size of any AFM tip and associated convolution of tip radius and nanowire width. Fig. 3c shows the width-to-height aspect ratio of the nanowires plotted as a function of the nanowire length. Nanowires shorter in length than ~120 nm exhibit ratios between 0.75 and 2.5 which we attribute to the seeding of the growth (gray area). Longer nanowires have an aspect ratio very close to unity: as the axial growth sets in, the nanowires appear to minimize surface area by maintaining a width to height ratio near unity (black markers). Fig. 3c includes also a significant number of long nanowires that form the basis of the transport measurements in the next section of this manuscript. The inset shows the dependence of width to height for the black markers of the main panel. The slope is 1.06.

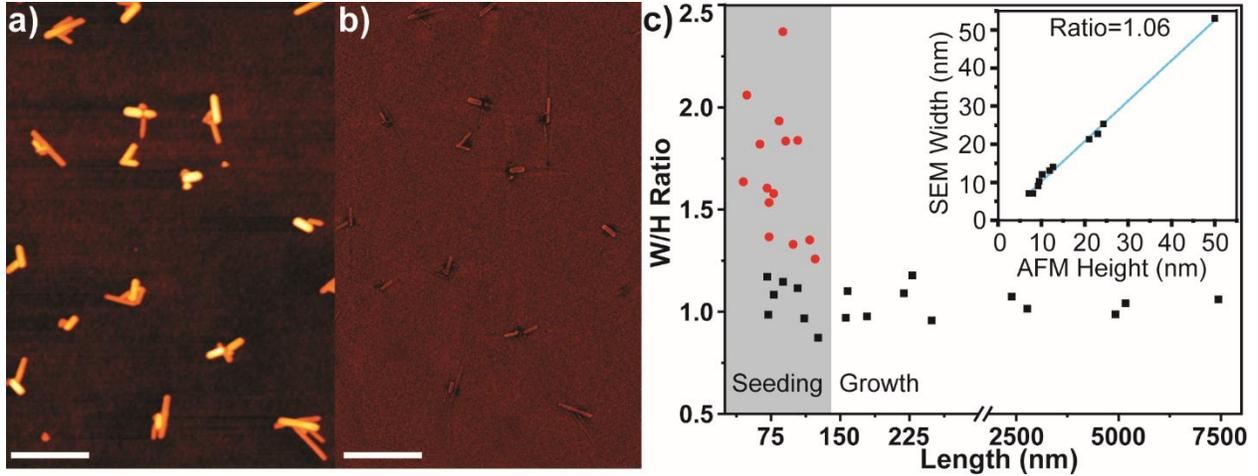

*Fig. 3 a,b) Atomic force microscopy (AFM) and scanning electron microscopy (SEM) images, respectively, of the same population of TaSe$_3$ 1D vdW nanowire bundles (scalebar is 500 nm); c) The width to height ratio of the 1D vdW TaSe$_3$ nanowires at seeding (grey area) and as uniaxial growth continues: longer wires have a width-to-height aspect ratio of practical unity. The inset plots SEM width vs. the AFM height of the wires indicated by black markers in the main panel.*

For measurement of the electrical transport properties of the TaSe$_3$ nanowires we employed electron beam lithography (EBL) to fabricate contacts consisting of 5 nm of yttrium for adhesion and 50nm of gold for conduction and stability. Previous studies of TaSe$_3$ nanowires[15-17] used encapsulation in *h*-BN to avoid surface decomposition by oxygen and moisture from the air. Striving to utilize only scalable methods in our work, we took a different approach: immediately following removal from the process tube, we cap the substrate containing the 1D vdW TaSe$_3$ nanowires with spin-coated polymethyl methacrylate (C5 PMMA) resist under a nitrogen-atmosphere. Subsequently, we characterize the nanowires by Raman spectroscopy through the PMMA film and then use the same resist film to fabricate electrical contacts. Care is taken to minimize the time between development and metal deposition so as to reduce air exposure of the sample. Immediately following metal liftoff, we again spin coat the sample with a layer of PMMA in a nitrogen glovebox. Subsequently, a second EBL process is performed to remove resist from the surface of the probe pads only. Electrical characterization proceeds in a nitrogen atmosphere.

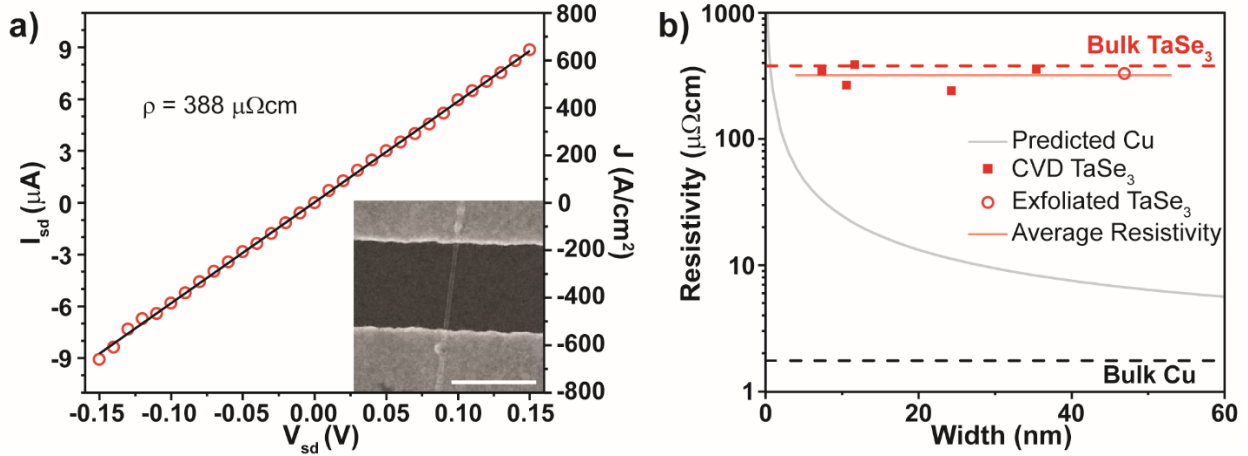

*Fig. 4 a) Source-Drain Current vs. Voltage ($I_{sd}$ vs. $V_{sd}$) for a 11.6 nm TaSe$_3$ nanowire; a linear dependence is observed in this two-probe measurement that corresponds to the current density shown on the y-axis and the resistivity highlighted. The inset shows an SEM image of a 1D vdW TaSe$_3$ nanowire bridging two electrodes (scalebar is 500 nm); b) resistivity of TaSe$_3$ nanowire bundles with width-to-height aspect ratios near unity (1.0-1.1) as a function of bundle width. For reference, we include bulk values for copper and TaSe$_3$ as dashed lines, as well as a prediction for the scaling of the copper resistivity with wire width based on Ref. [21]*

Fig. 4a shows a typical current-voltage (I-V) diagram measured on a TaSe$_3$ nanowire with width and height of ~11.6 nm each. We use electrodes separated by 500 nm (inset). A linear I-V dependence is observed, from which a resistivity ρ of 388 μΩcm is obtained. The secondary y-axis of Fig. 4a shows the corresponding current density *J*.

We tested a large number of nanowires, each time followed by AFM and SEM characterization of their respective width and height. Fig. 4b shows the resistivity of wires with width ≤ 50 nm and cross section aspect ratios near unity (1.0-1.1). In a few cases we found wires that exhibited significant non-linear response for small voltage, which we ascribe to contact resistance. These were omitted from Fig. 4b. The figure reveals that the specific resistance of 1D vdW TasSe$_3$ nanowires is independent of the cross section of the wire bundle down to 7 nm in width and height. This presents a marked contrast to the behavior of copper at the nanoscale for which Fig. 4b includes a reference line based on the work of Steinhogel et al. *[21]* assuming surface *p* and grain boundary *R* scattering amplitudes of 0.5 and 0.6, respectively.

If scattering does not limit the scaling of the conductivity of TaSe$_3$ nanowires, we explore whether quantum confinement perpendicular to the nanowire direction may do so. To this end, we assume that the conductivity of a metallic material in first order scales with its density of state (DOS) near the Fermi level ($E_F$) and calculate the DOS of various nanowire bundle configurations. We employ density functional theory as implemented in the Vienna Ab initio Simulation Package (VASP)[33] using the projector augmented wave method[34, 35] and treating the electron exchange-correlation interaction by the generalized gradient approximation (GGA) functional of Perdew, Burke, and Ernzerhof (PBE).[36] All calculations use periodic boundary conditions and the Brillouin zone was sampled by a 2×7×2 Monkhorst−Pack k-point grid.[37]

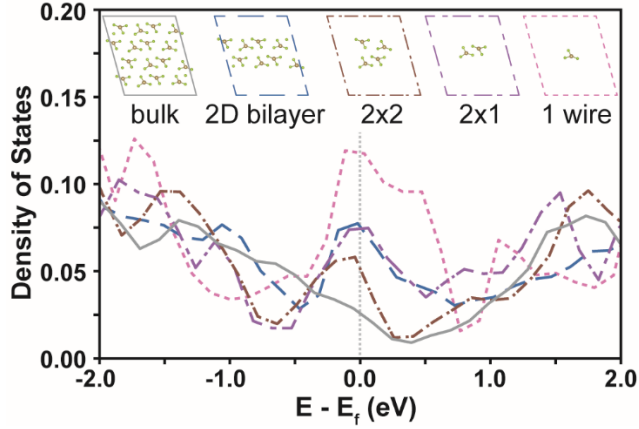

*Fig. 5* *Calculated density of states (DOS) near the Fermi level ($E_F$) for bulk TaSe$_3$, a 2D bilayer of wires, a 2×2 wire bundle, a 2×1 wire bundle and a single wire. The confined wire geometries have higher DOS at $E_F$ than the bulk material suggesting that quantum confinement does not adversely affect charge transport in TaSe$_3$.*

Fig. 5 compares the density of states (DOS) of an infinite bulk of TaSe$_3$ (solid line) to an infinite 2D bilayer of wires, bundles of 2×2 and 2×1 wires, as well as a single wire. In each case we find considerable DOS near the Fermi level ($E_F$) and no band gap. Indeed, as the wire bundle is thinned to a bilayer and a finite number of wire stacks, the DOS near $E_F$ increases. This finding suggests that – at least for some wire bundle geometries – a higher native conductivity is possible than for the bulk case. We recognize, however, that this analysis omits fundamental stability limitations (such as the Mermin-Wagner-theorem[38]) yet we note that additional research is necessary to fathom their impact as shown for graphene.[39] The smallest wire bundle for which we obtained transport measurements had a width and height of ~ 7nm (Fig. 4b) and thus as few as ~10x10 wires in parallel; it is almost an order of magnitude larger than the computationally readily tractable ones of Fig. 5.

Finally, we turn to the stability of the 1D vdW nanowires with regards to degradation under transport. Measurements of the low-frequency noise are commonly used to assess the quality and reliability of conventional[40-44] and novel 2D materials [16, 18, 45] for device applications. Changes in the noise spectra can serve as a convenient indicator of the onset of electromigration and other material degradation mechanisms. In the context of interconnect research, the low-frequency noise can provide a fast estimate of the device's mean time to failure. The low-frequency noise measurements were performed using an experimental setup consisting of a "quiet" battery, a potentiometer biasing circuit, a low noise amplifier, and a spectrum analyzer; additional details have been reported elsewhere.[46, 47]

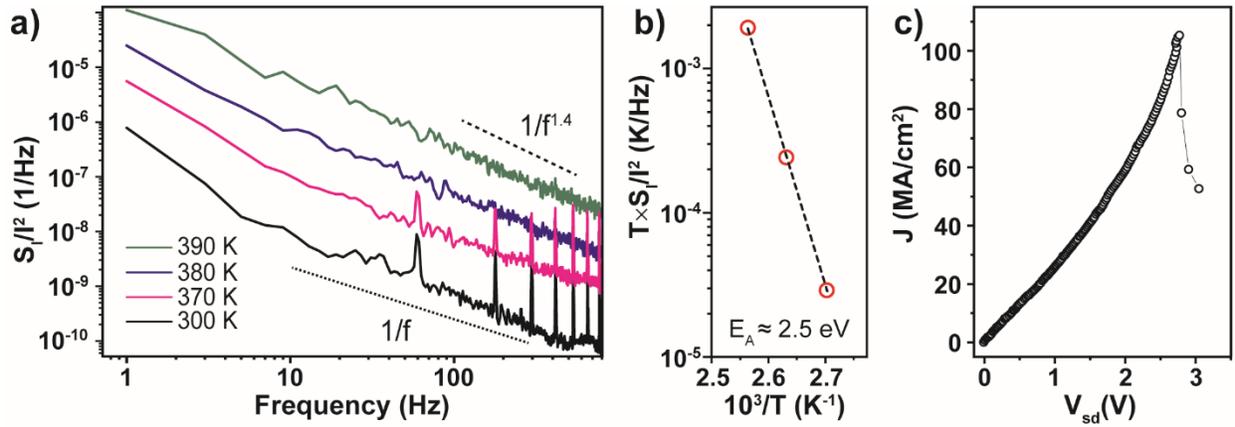

*Fig. 6*: a) Normalized current noise spectral density of a ~10×10 nm$^2$ (width × height) CVD 1D vdW TaSe$_3$ nanowire bundle at different temperatures T using a source-drain voltage $V_{sd}$ of 0.1 V. b) Arrhenius plot of T×S$_I$/I$^2$ vs. 1000/T for a frequency f of 11 Hz. The extracted activation energy $E_A$ is 2.57 eV. c) Current density J response of a ~7×7 nm$^2$ nanowire as the voltage is slowly increased. Failure occurs at a current density in excess of 10$^8$ A/cm$^2$.

Figure 6a shows the normalized current noise spectral density $S_I/I^2$ as a function of frequency $f$ for a TaSe$_3$ nanowire with a cross-section of ~ $10\times10$ nm$^2$. The noise level $S_I/I^2$ of ~$10^{-8}$ Hz$^{-1}$ at $f=10$ Hz and $T=300$K in the downscaled CVD TaSe$_3$ nanowires is appreciably low. Although it is higher than measured in conventional metals[48, 49] and also for larger cross section exfoliated TaSe$_3$ nanowires,[16, 49] the latter is expected because the noise originating from a volume of independent fluctuators scales inversely proportional to the size of the volume.[40]

Fig. 6a reveals increase in the noise level with increasing temperature. The informative frequency range from 10 Hz to ~400 Hz exhibits a deviation from pure $1/f$ noise; approximately $1/f^{1.4}$ provides the best fit at elevated temperatures. This trend is consistent with observations on thicker exfoliated TaSe$_3$ nanowires.[16] In metals the deviation from $1/f$-type behavior is commonly attributed to the onset of electromigration. We construct an Arrhenius plot of $T\times S_I/I^2$ vs. 1000/T (Fig. 6b) to extract the activation energy ($E_A$) for the noise inducing process in CVD 1D vdW TaSe$_3$ nanowires. The resultant value of ~2.5 eV is larger than that for exfoliated TaSe$_3$ nanowires[16] and more than twice that for electromigration in copper (0.76-1.10 eV) and aluminum (0.67- 1.14 eV) using similar measurements.[43, 44, 49, 50] This finding suggests very good resistance of CVD TaSe$_3$ 1D vdW nanowire to electromigration.

The electromigration resilience of the CVD 1D vdW TaSe$_3$ nanowires is confirmed by successively increasing the voltage applied to a $7\times7$ nm$^2$ wire bundle (about 100 parallel Ta-Se$_3$ stacks in total). Fig. 6c shows that the wire bundle was able to sustain in excess of $10^8$ A/cm$^2$ before electrical breakdown. This value is an order of magnitude higher than that found for exfoliated TaSe$_3$, and also slightly better than our previous findings for ZrTe$_3$.[18] We note that acquiring the dataset of Fig. 6c took almost an hour of slowly increasing the bias, so that self-heating may have contributed to reduced overall current carrying capacity and increased susceptibility to electromigration. Embedding the nanowire with a better thermal sink than the underlying $SiO_2$ substrate and the surrounding PMMA resist may result in even higher sustained current densities.

In conclusion, we find that chemical vapor deposition allows the growth of $TaSe_3$ nanowires on a $SiO_2$ substrate that are competitive in resistivity to conventional metals scaled to sub-10-nanometer wire diameters, while offering significantly enhanced electromigration resilience and breakdown current.

**Acknowledgements:**

This project was supported by the Semiconductor Research Corporation (SRC) under contract 2018-NM-2796. The noise measurements and the computational work were supported, in part, by the National Science Foundation (NSF) under grants EFRI-1433395 and DMR-1455050, respectively. Ancillary support originated form the US Air Force Office of Scientific Research under Grant FA9550-14-1-0378.

**Supporting Material:**

Supporting material is available showing electron microscopy imaging of the CVD precursors and thermal behavior of the resistance of $TaSe_3$ nanowires.